\documentclass[reprint,amssymb, amsmath, aps, superscriptaddress, showpacs, footinbib, longbibliography, prl]{revtex4-1}
\usepackage{amssymb}
\usepackage{amsmath}
\usepackage{xcolor}
\usepackage{graphicx,hyperref}
\usepackage{footmisc,booktabs}
\usepackage{gensymb}
\usepackage{sidecap}
\usepackage{microtype}

\begin{document}

\title{Anisotropic field-induced ordering in the triangular-lattice quantum spin liquid NaYbSe$_2$}

\author{K.~M.~Ranjith}
\email{ranjith.kumar@cpfs.mpg.de}
\affiliation{Max Planck Institute for Chemical Physics of Solids, 01187 Dresden, Germany}
\author{S.~Luther}
\affiliation{Hochfeld-Magnetlabor Dresden (HLD-EMFL) and W\"{u}rzburg-Dresden Cluster of Excellence ct.qmat,
      Helmholtz-Zentrum Dresden-Rossendorf, 01328 Dresden, Germany}
\affiliation{Institut f\"{u}r Festk\"orper- und Materialphysik, TU Dresden, 01062 Dresden Germany}
\author{T.~Reimann}
\affiliation{Hochfeld-Magnetlabor Dresden (HLD-EMFL) and W\"{u}rzburg-Dresden Cluster of Excellence ct.qmat,
      Helmholtz-Zentrum Dresden-Rossendorf, 01328 Dresden, Germany}
\author{B.~Schmidt}
\affiliation{Max Planck Institute for Chemical Physics of Solids, 01187 Dresden, Germany}
\author{Ph.~Schlender}
\affiliation{Faculty of Chemistry and Food Chemistry, TU Dresden, 01062 Dresden, Germany}
\author{J. Sichelschmidt}
\affiliation{Max Planck Institute for Chemical Physics of Solids, 01187 Dresden, Germany}
\author{H.~Yasuoka}
\affiliation{Max Planck Institute for Chemical Physics of Solids, 01187 Dresden, Germany}
\author{A.~M.~Strydom}
\affiliation{Highly Correlated Matter Research Group, Department of Physics, University of
Johannesburg, PO Box 524, Auckland Park 2006, South Africa}
\author{Y. Skourski}
\affiliation{Hochfeld-Magnetlabor Dresden (HLD-EMFL) and W\"{u}rzburg-Dresden Cluster of Excellence ct.qmat,
      Helmholtz-Zentrum Dresden-Rossendorf, 01328 Dresden, Germany}
\author{J.~Wosnitza}
\affiliation{Hochfeld-Magnetlabor Dresden (HLD-EMFL) and W\"{u}rzburg-Dresden Cluster of Excellence ct.qmat,
      Helmholtz-Zentrum Dresden-Rossendorf, 01328 Dresden, Germany}
\affiliation{Institut f\"{u}r Festk\"orper- und Materialphysik, TU Dresden, 01062 Dresden Germany}
\author{ H.~K\"{u}hne}
\affiliation{Hochfeld-Magnetlabor Dresden (HLD-EMFL) and W\"{u}rzburg-Dresden Cluster of Excellence ct.qmat,
      Helmholtz-Zentrum Dresden-Rossendorf, 01328 Dresden, Germany}
\author{Th.~Doert}
\affiliation{Faculty of Chemistry and Food Chemistry, TU Dresden, 01062 Dresden, Germany}
\author{M. Baenitz}
\email{michael.baenitz@cpfs.mpg.de}
\affiliation{Max Planck Institute for Chemical Physics of Solids, 01187 Dresden, Germany}

\date{\today}

\begin{abstract}\noindent

High-quality single crystals of NaYbSe$_2$, which resembles a perfect triangular-lattice antiferromagnet without the intrinsic disorder, are investigated by magnetization and specific-heat, as well as the local probe techniques nuclear magnetic resonance (NMR) and electron spin resonance (ESR). The low-field measurements confirm the absence of any spin freezing or long-range magnetic order down to 50~mK, which suggests a quantum spin liquid ground (QSL) state with gapless excitations. Instability of the QSL state is observed upon applying magnetic fields. For the $H\bot c$ direction, a field-induced magnetic phase transition is  observed above 2~T from the $C_{\rm p}(T)$ data, agreeing with a clear $\frac{M_s}{3}$ plateau of $M(H)$, which is associated with an up-up-down (uud) spin arrangement. For the $H\|c$ direction, a field-induced transition could be evidenced at a much higher field range (9 - 21~T). The $^{23}$Na NMR measurements provide microscopic evidence for field-induced ordering for both directions. A reentrant behaviour of $T_{\rm N}$, originating from the thermal and quantum spin fluctuations, is observed for both directions. The anisotropic exchange interactions $J_{\perp}\simeq$ 4.7~K and $J_z\simeq$2.33~K are extracted from the modified bond-dependent XXZ model for the spin-$\frac{1}{2}$ triangular-lattice antiferromagnet. The absence of magnetic long-range order at zero fields is assigned to the effect of strong bond-frustration, arising  from the complex spin-orbit entangled $4f$ ground state. Finally, we derive the highly anisotropic magnetic phase diagram, which is discussed in comparison with the existing theoretical models for spin-$\frac{1}{2}$ triangular-lattice antiferromagnets.
\end{abstract}

\maketitle

The search for a quantum spin liquid (QSL) state, a highly entangled state with fractionalized excitations, has been at the forefront of current condensed-matter physics research~\cite{balents2010,Zhou2017,Wen2019}. Spin-$\frac{1}{2}$ triangular-lattice antiferromagnets (TLAFs) with antiferromagnetic (AF) nearest-neighbor (NN) interactions are considered as the prime candidate for this search, in which Anderson proposed a QSL ground state more than 40 years ago~\cite{anderson1973}. Even though, the ground state of the Heisenberg TLAF model is now known to be a 120~$^{\circ}$ AF ordered state~\cite{bernu1994,capriotti1999}, it can be easily perturbed  by different mechanisms. For instance, the presence of a next-NN interaction~\cite{Zhu2015,Iqbal2016},  anisotropic planar NN exchange interactions~\cite{Yunoki2006}, or spatially random exchange interactions (bond-randomness)~\cite{Shimokawa2015} may  suppress the ordering and consequently stabilize the QSL ground states. Experimentally, several QSL candidate materials have since been reported including $\kappa$-(BEDT-TTF)$_2$Cu$_2$(CN)$_3$, EtMe$_3$Sb[Pd(dmit)$_2$]$_2$, and Ba$_3$CuSb$_2$O$_9$~\cite{Shimizu2003,Yamashita2008,yamashita2010,Isono2016,Yamashita2011,itou2008,Zhou2011}.

 Rare-earth based frustrated spin systems have recently gained much attention~\cite{Rau2018,Maksimov2019}. The strong spin-orbit coupling (SOC) associated with the rare-earth ion leads to highly anisotropic and bond-dependent exchange interactions between the moments and may host strong quantum fluctuations, which can play a crucial role in stabilising the QSL state~\cite{chen2009,jackeli2009,ishizuka2014,lu2017}. YbMgGaO$_4$, a 4$f$ based triangular-lattice system, has been investigated rigorously and proposed as a promising QSL candidate~\cite{li2015,li2015a}. The absence of magnetic ordering down to 48~mK, an anomalous $C_v\propto T^{2/3}$ behaviour~\cite{li2015}, and the presence of an excitation continuum~\cite{shen2016,paddison2017} are attributed to the QSL state with spinon excitations. On the other hand,  the lack of a significant contribution from the
magnetic excitations  to the thermal conductivity~\cite{Xu2016}, a persistent excitation continuum even in the high-field polarized phase~\cite{paddison2017}, and the sizeable broadening of crystal electric field (CEF) excitations~\cite{li2017b,paddison2017} question the QSL ground state. The inherent structural disorder, a random distribution of Mg$^{2+}$ and Ga$^{3+}$ ions present in this system  mimic a spin-liquid-like state at low temperatures, instead a robust collinear/stripe magnetic order is expected for a disorder-free version of YbMgGaO$_4$, theoretically~\cite{zhu2017}.

On the other hand, theoretical studies propose a fascinating phase diagram for TLAFs in external magnetic fields~\cite{chubokov1991,Kawamura1985,schmidt:17}. Thermal and quantum spin fluctuations can  favor different coplanar (an oblique
version of the 120$^{\circ}$ state, $Y$ phase and 2:1 canted phase) and collinear (up-up-down,uud phase) spin configurations. The stabilization of the collinear uud phase results in a $\frac{M_s}{3}$ plateau ($M_s$ is the saturation magnetization) for a finite field range in the magnetization curve~\cite{chubokov1991,Kawamura1985}. Experimentally,  such a phase diagram has been observed in the spin-$\frac{1}{2}$  triangular-lattice compounds  Cs$_2$CuBr$_4$~\cite{Ono2003,Alicea2009} and Ba$_3$CoSb$_2$O$_9$ ~\cite{Fortune2009,Susuki2013}.

The family of Yb-dichalcogenide delafossites NaYb$Ch_2$ ($Ch$ = O, S, and Se) has recently been explored as a perfect spin-$\frac{1}{2}$  TLAF without inherent disorder~\cite{liu2018,baenitz2018,Ranjith2019}. Here, the combination of the SOC and the crystal electric field (CEF) leads to a Kramers’ doublet ground state for the Yb$^{3+}$ ion, so that the low temperature properties can be described by an effective spin-$\frac{1}{2}$ Hamiltonian. The energy gap between the ground state and the first excited doublet was found to be $\Delta/k_{\rm B}\sim$ 200 and $\sim$400~K for NaYbS$_2$~\cite{baenitz2018} and  NaYbO$_2$~\cite{Ranjith2019,Ding2019}, respectively, which suggests an effective spin-$\frac{1}{2}$ ground state. Experimentally, this is confirmed by specific heat measurements, which reveal a magnetic entropy of Rln(2) per Yb$^{3+}$ ion~\cite{baenitz2018,Ranjith2019}. Zero-field $\mu$SR and heat capacity measurements down to 50~mK confirmed the absence of magnetic long-range order (LRO), and suggested a QSL ground state with gapless excitations for NaYbS$_2$ and NaYbO$_2$~\cite{baenitz2018,Ranjith2019,Ding2019}. Furthermore, an excitation continuum with the spectral weight accumulating around the $K$-point, which is expected for a QSL state, was observed in neutron studies of polycrystalline NaYbO$_2$~\cite{Ding2019}. In NaYbO$_2$, this QSL state is found to be suppressed by the application of external magnetic fields, and a field-induced ordered state was evidenced by magnetization, heat capacity, $^{23}$Na NMR ~\cite{Ranjith2019}, and neutron diffraction~\cite{Bordelon2019}.
NaYb$Ch_2$ ($Ch$ = O, S, and Se) provide a unique platform to study the field-induced crossover from an emerging spin-orbit driven QSL at low fields to an isotropic 2D planar spin-$\frac{1}{2}$  TLAF  with particular types of LRO at much higher fields.

It should be mentioned that, in contrast to previous studies on   NaYbO$_2$~\cite{Ranjith2019,Ding2019,Bordelon2019},  NaYbSe$_2$ is available in single crystalline form allowing for the determination of anisotropic static and dynamic magnetic properties. Here, we show that the emergent QSL ground state is associated with the quantum fluctuations and strong frustration , arising from the complex bond-dependent exchange interactions. The triangular arrangement of edge-shared YbSe$_6$ octahedra favors bond-dependent frustration. Each bond can be represented by an individual exchange matrix where the off-diagonal terms are responsible for the strong bond frustration. As these off-diagonal terms are usually small, they can be wiped out upon the application of external fields, converting the system into a more classical anisotropic planar  spin-$\frac{1}{2}$ TLAF.

\section*{Experimental}
Single crystals of NaYbSe$_2$ are synthesized by using NaCl, Yb metal grains, and Se as starting materials. All chemicals are stored and handled in an argon-filled glove-box. For the syntheses of NaYbSe$_2$, 825.5 mg (20 mmol, 20 eq.) NaCl (lab stock, dried), 122.2 mg (0.71 mmol, 1 eq) Yb (MaTeck, 99.92\%, rod) and 133.9 mg (1.7 mmol, 2.4 eq) Se (lab stock, sublimed) are mixed into a fused silica tube with internal glassy carbon crucible. The tube is evacuated and flame sealed, placed upright in a muffle furnace and heated up to 400~$^{\circ}$C with 180~K/h and equilibrated there for 2 hours, before heated to 850~$^{\circ}$C with 20~K/h. After one week, the temperature is decreased with 40~K/h to room temperature. After opening at ambient atmosphere, the sample is removed and the alkali halide is rinsed off with water. The NaYbSe$_2$ crystals are washed several times with water and ethanol and dried in the hood.
Polycrystalline samples of NaLuO$_2$ for comparison were synthesized by a solid-state reaction starting from Na$_2$O and Lu$_2$O$_3$. The mixed powder was ground in a mortar under Ar atmosphere, placed in an alumina crucible and heated up to 800~$^{\circ}$C for 12 hours in air. The material was then ground again and washed with distilled water and ethanol to remove the residual Na$_2$O. The phase purity of the samples was confirmed by powder x-ray diffraction (XRD) with a Stoe Stadi-P powder diffractometer using Cu-K$\alpha$  radiation ($\lambda$ = 1.54056 nm) and a Mythen 1K strip detector in flat-sample transmission mode. In the diffraction pattern, all reflections can be assigned to the designated target phases; no reflections of by-products are observed. Due to the thin plate-like shape of the crystals  and the respective preferred orientation of the flakes, reflections with major contribution of Miller index $l$ are highly overestimated.

Magnetization measurements were performed with a superconducting quantum interference device (SQUID) magnetometer [magnetic property measurement system (MPMS)] and a vibrating sample magnetometer (VSM). The MPMS with a $^3$He insert was used for the low temperature magnetization (down to 500 mK). The angular-resolved susceptibility measurements were performed using a commercial horizontal rotator from Quantum Design. Specific-heat measurements were conducted with a commercial PPMS (physical property measurement system, Quantum Design) with a $^3$He insert down to 350~mK on NaYbSe$_2$ single crystals and pressed pellets of NaLuO$_2$ powder samples. The specific heat down to 50~mK was obtained in a $^3$He/$^4$He dilution fridge setup. The magnetization measurements in pulsed magnetic fields up to 35 T were performed at the Dresden High Magnetic Field Laboratory (HLD) with a compensated pickup-coil system in a pulse field magnetometer in a home-build $^3$He cryostat.

Electron spin resonance (ESR) experiments were carried out with a standard continuous-wave spectrometer. We measured the power $P$ absorbed by the sample from a transverse magnetic microwave field (X-band, $\nu $ = 9.4~GHz) as a function of an external, static magnetic field $\mu_0H=B$. In order to improve the signal-to-noise ratio, a lock-in technique was used, which yields the derivative of the resonance signal dP/dH. Nuclear magnetic resonance (NMR) measurements were performed using  Tecmag Apollo and Redstone spectrometers with a standard probe and a sweepable superconducting magnet. Field sweep NMR spectra were obtained by the integration of spin echo signals at a fixed frequency.  The spectra at low temperatures were constructed from frequency swept Fourier transform sums.   Spin-lattice relaxation rates were measured using a standard inversion recovery method, where the nuclear magnetization $M(t)$ was obtained from the recovery of the spin-echo magnitude as a function of the time interval $\tau$ between the inversion pulse ($\pi$ pulse) and the $\pi/2 -\pi$ spin-echo sequence.

\section*{Results}
\subsection{Crystal structure and magnetic anisotropy:}
\begin{figure}
  \centering
  \includegraphics[clip,width=1\columnwidth]{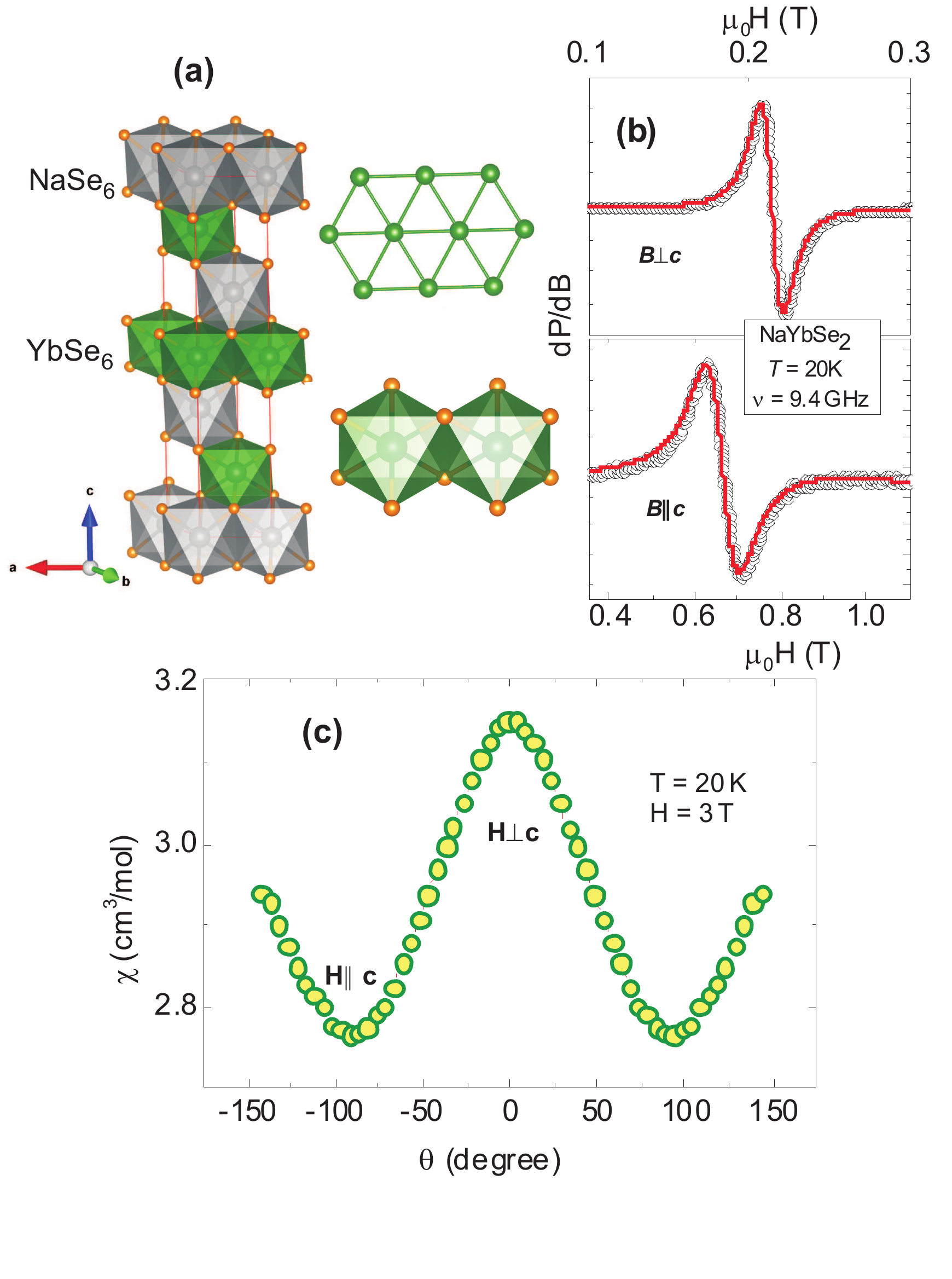}
  \caption{(a) Schematic crystal structure of NaYbSe$_2$. Yb-triangular layers (green) are formed by edge-shared YbSe$_6$ octahedra in the $ab$-plane. (b) ESR spectra measured at 20~K for both the $H\bot c$ and $H\| c$ directions. (c) Angular dependence of  the magnetic susceptibility obtained at 3~T and 20~K for a NaYbSe$_2$ single crystal.}\label{fig1}
\end{figure}
Single crystals of NaYbSe$_2$ with lateral crystallite size in the millimeter range were synthesized by flux reactions~(see the Supplementary Information ~\footnote{see the Supplemental Material which include Ref.-~\cite{fulde:79,moriya:60,li2015a,Rau2018,zhu2018,steppke2010,itou2010,shockley2015,ranjith2018,Ranjith2019}}). NaYbSe$_2$ has a delafossite-type structure with $R\bar{3}m$ space group (No:166). As shown in Fig.~\ref{fig1}(a), the crystal structure is composed of edge-shared YbSe$_6$ and NaSe$_6$ octahedra. The YbSe$_6$  octahedra are weakly distorted with two “large” equilateral triangles with edge length $l\simeq$ 4.06~\AA~ and six “small” isosceles triangles with two edges of length $s\simeq$ 3.7\AA~ and one edge of length $l$. The distorted octahedra are tilted along the Yb-Yb bond by an angle $\alpha$ = cos$^{-1}(l/\sqrt{3}s)\simeq$ 53.33$^\circ$, such that the large triangular faces are perpendicular to the crystallographic c direction. The tilt of the YbSe$_6$ octahedra in NaYbSe$_2$ is smaller compared to that in NaYbO$_2$ ($\alpha\simeq 47.70^\circ$) and NaYbS$_2$ ($\alpha\simeq 48.60^\circ$) and the tilting angle $\alpha\sim54.74^{\circ}$ for an undistorted octahedron. The NaSe$_6$ octahedra are distorted as well. The shortest Yb-Yb distances are found in the $ab$ plane, forming triangular layers which are well separated by edge-shared NaSe$_2$ layers. A rhombohedral stacking (ABCA stacking) of magnetic Yb-triangular layers provides additional strong frustration of the interlayer interactions.

Figure~\ref{fig1}(c) shows the angle-resolved magnetic susceptibility  $\chi(\theta)$ of a single crystal sample where the crystal was successively rotated from $ab$ plane ($\theta = 0$~degree) to $c$-direction ($\theta = \pm90$~degree).  The presence of well defined minima and maxima clearly reveals the existence of a significant easy-plane type anisotropy in NaYbSe$_2$.  The anisotropy of $\chi$ vanishes towards high temperatures. The susceptibility anisotropy of $\chi_{\perp}/\chi_{\parallel}\simeq$ 1.5 for NaYbSe$_2$ at $T\simeq$ 20~K  is smaller than that of NaYbS$_2$ ($\sim$2) at the same temperature. Usually the anisotropy in $\chi$ captures both spin and exchange anisotropy.

 Electron spin resonance (ESR) provides direct information about the spin-anisotropy. Figure~\ref{fig1}(b) shows the ESR lines at 20~K for the field applied along the $ab$-plane and $c$ axis. The well resolved and narrow ESR spectra obtained for NaYbSe$_2$ single crystals indicate the absence of intrinsic structural disorder. This is in contrast to the recently reported Yb-based QSL candidate YbMgGaO$_4$, for which the structurally distorted occupancy of Ga and Mg leads to an off-center displacement of Yb$^{3+}$ ions, resulting in a distribution of $g$ values with rather broad ESR spectra. Furthermore, the ESR measurements for $H\perp c$ [Fig.~\ref{fig1}(b) bottom panel] and $H\parallel c$ [Fig.~\ref{fig1}(b) top panel] of  NaYbSe$_2$ yield strongly anisotropic $g$-values of $g_{\bot}$ = 3.13(4) and $g_{\|}$ = 1.01(1). The $g$-value anisotropy, which is corresponding to the spin anisotropy, is obtained as $g_{\bot}/g_{\|}\simeq$ 3.1, which is significantly smaller than that of NaYbS$_2$ ($g_{\bot}/g_{\|}\simeq$ 5.6)~\cite{baenitz2018,Joerg2019a}. Due to the lack of neutron scattering data, a first estimate of the energy gap, $\Delta$ between the ground state doublet and the first excited doublet could be obtained from the temperature dependence of the ESR line width, which increases exponentially towards high temperatures due to an Orbach process. The obtained value of $\Delta/k_{\rm B}\sim$ 160$\pm$30~K confirms the low temperature pseudospin-$\frac{1}{2}$ ground state~\cite{Joerg2019}.
\subsection{Low field thermodynamics:}

\subsubsection{Magnetic susceptibility}

\begin{figure}
  \centering
  \includegraphics[clip,width=1\columnwidth]{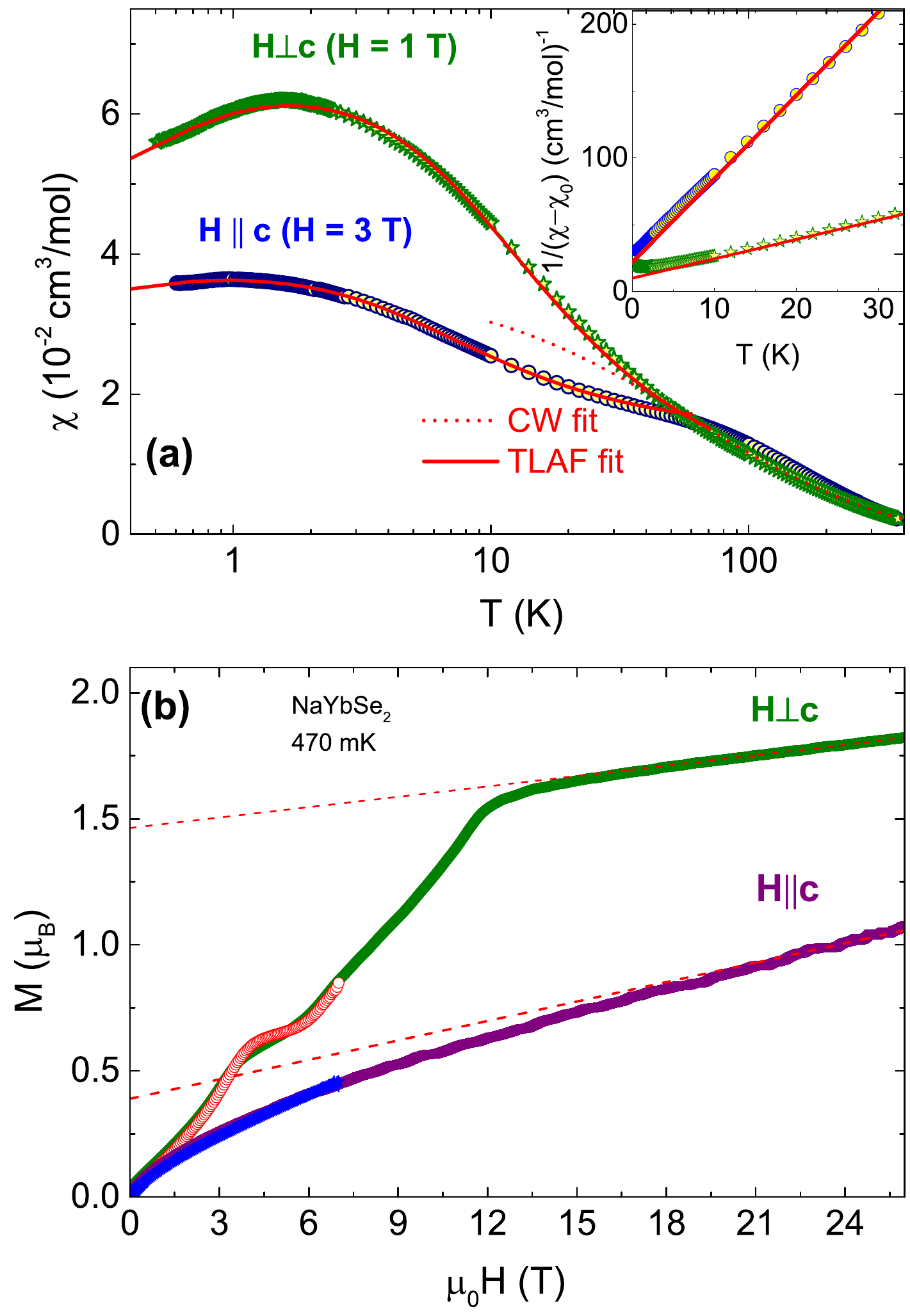}
 \caption{(a) Magnetic susceptibilities ($\chi_{\|}$ and $\chi_{\bot}$) of NaYbSe$_2$ single crystals. Red dotted and  solid lines correspond to the high temperature Curie-Weiss (CW) fit and TLAF fitting (see text), respectively. The inset shows the low temperature inverse magnetic susceptibility after subtracting $\chi_{VV}$ along with the CW fit for both directions. (b) Isothermal magnetization $M(H)$ measured at 470~mK for both directions. Dashed lines represent the van Vleck contribution (see text).} \label{fig2}
\end{figure}
Figure~\ref{fig2}(a) shows the temperature dependence of the magnetic susceptibility of NaYbSe$_2$ measured for fields applied in the $ab$ plane, $\chi_\bot(T)$ and along the $c$ axis, $\chi_\|(T)$. At higher temperatures $T\geq70\,\rm K$, $\chi(T)$ does not show a significant anisotropy, and can be well described by
\begin{equation}
	\left.\chi(T)\right|_{T\geq70\,\text K}
	=
	\chi^\text{dia}+\chi^\text{VV}
	+\frac{N_\text L\mu_0}3\frac{\mu_\text{eff}^2}{T-\Theta_\text{CW}}
\end{equation}
where $\chi^\text{dia}$ is a temperature-independent core diamagnetic contribution and $\chi^\text{VV}$ is the van Vleck susceptibility in the limit $\mu_0H\to0$ neglecting the Curie term,
\begin{equation}
\begin{aligned}
	\chi^\text{VV}
	&=
	N_\text L\mu_0\left(g_j\mu_\text B\right)^2
	\sum_{n}\sum_{m\ne n}
	M_{nm}^0\frac{p_n^0-p_m^0}{E_m^0-E_n^0},
	\\
	M_{nm}^0
	&:=
	\sum_{\alpha\alpha'}\left|\left\langle n^0,\alpha\left|
	J_a\right|m^0,\alpha'\right\rangle\right|^2.
\end{aligned}
\label{eqn:chivv}
\end{equation}
with the usual meaning of the constants, $g_j=8/7$ denotes the Land\'{e} $g$ factor for the Yb ion. Latin indices $m$ and $n$ indicate sums over the four CEF doublets with energies $E_m^0$ and $E_n^0$, Greek indices indicate the summation over the degenerate states within a doublet. The thermal population of a doublet $n$ is given by $p_n^0=\exp\left[-E_n^0/(k_\text BT)\right]/Z$, where $Z=2\sum_n\exp\left[-E_n^0/(k_\text BT)\right]$ denotes the partition function of the CEF doublets. The operator $J_a$ is the component of the total angular momentum operator in direction of the applied field, the bra and ket vectors denote the CEF eigenstates.

We obtain $\chi^\text{dia}+\chi^\text{VV}\approx2\times10^{-4}\,{\rm cm}^3/{\rm mol}$, a Weiss temperature $\Theta_\text{CW}=-66\,\rm K$, and an effective moment $\mu_\text{eff}=4.5\,\mu_\text B$, which are in good agreement with the literature~\cite{liu2018}. The value of $\mu_{\rm eff}/\mu_\text B$ is close to the free-ion value $p=g_j\sqrt{j(j+1)}\approx4.54$ for Yb$^{3+}$ ($j=\frac{7}{2}$).

A deviation from the Curie-Weiss (CW) behaviour together with a large anisotropy is observed at low temperatures, indicating the evolution of a ground-state CEF doublet with small exchange couplings. $\chi_\perp(T)$ and $\chi_\parallel(T)$ did not show any signatures of magnetic LRO down to $0.5\,\rm K$. Instead, $\chi_\bot(T)$ reveals a pronounced broad maximum centered around $2\,\rm K$, a characteristic feature of strong magnetic short-range correlations associated with the low-dimensionality of the magnetic exchange. For $H\parallel c$, a similar broad maximum is observed at a slightly lower temperature.

Below 70~K, the magnetic properties of NaYbSe$_2$ are determined by the lowest Kramers doublet and can be described by an effective spin-$\frac{1}{2}$ model similar to NaYbO$_2$ and NaYbS$_2$~\cite{Ranjith2019,baenitz2018}. At low temperatures and finite applied magnetic fields, $\chi(T,H)$ has a sizeable contribution from the van Vleck susceptibility $\chi^{\rm VV}(T,H)$ which arises from the excitations across the Zeeman-split CEF levels. In order to determine the $\chi^{\rm VV}$ contribution experimentally, we have measured the isothermal high-field magnetization $M(H)$ at $T=470\,\rm mK$ for both the directions. As seen in Fig.~\ref{fig2}(d), the in-plane magnetization (with $H\perp c$) has a kink at $\mu_0H_\bot^s\approx12\,\rm T$ and increases linearly with higher fields. The linear behaviour  of $M_\perp(H)$ above about $13\,\rm T$ is due to the van Vleck contribution, which can be estimated as $\chi^{\rm VV}_{\perp}\simeq0.0137\,\mu_{\rm B}/{\rm T}\simeq7.8\times 10^{-4} \,{\rm cm}^3/\rm mol$.

The van Vleck contribution is found to be much larger for $H\parallel c$, but it is difficult to determine, as full saturation is not reached for $\mu_0H\leq22\,\rm T$. $M_\parallel(H)$ is expected to saturate at much higher fields compared to $M_\perp(H)$ (see below). A linear fitting above $\mu_0H=24\,\rm T$ yields $\chi^{\rm VV}_{\parallel}\simeq0.0257\,\mu_{\rm B}/{\rm T}\simeq147\times 10^{-4}\,{\rm cm}^3/{\rm mol}$. The obtained saturation magnetizations $M^\text s_{\parallel}\simeq0.49\,\mu_{\rm B}$ and $M^\text s_{\perp}\simeq1.5\,\mu_{\rm B}$ are in good agreement with the expected values $M^\text s_{\perp,\parallel}=(1/2)g_{\perp,\parallel}\mu_{\rm B}$ ($\sim$0.5 and $\sim$1.5~$\mu_{\rm B}$ for $H\parallel c$ and $H\perp c$, respectively).

After subtracting $\chi^{\rm VV}$, $\chi(T)$ for $T\leq30\,\rm K$ could be fitted with a Curie-Weiss law [see inset of Fig.~\ref{fig2}(a)], which yields $\Theta_{\rm CW}^{\perp}\simeq-7\,\rm K$, $\mu_{\rm eff}^{\perp}\simeq2.43\,\mu_{\rm B}$ and $\Theta_{\rm CW}^{\|}\simeq-3.5\,\rm K$, $\mu_{\rm eff}^{\parallel}\simeq1.1\,\mu_{\rm B}$ for the $ab$-plane and $c$-direction, respectively. The effective moments are in good agreement with $\mu_{\perp,\parallel}=g_{\perp,\parallel}\sqrt{S(S+1)}\mu_{\rm B}$ for a pseudospin-$\frac{1}{2}$ ground state with the $g$ values obtained from our ESR measurements (2.7~$\mu_{\rm B}$ for $H\perp c$ and 0.9~$\mu_{\rm B}$ for $H\parallel c$).

\subsubsection{Specific heat}
\begin{figure}
  \centering
  \includegraphics[clip,width=1\columnwidth]{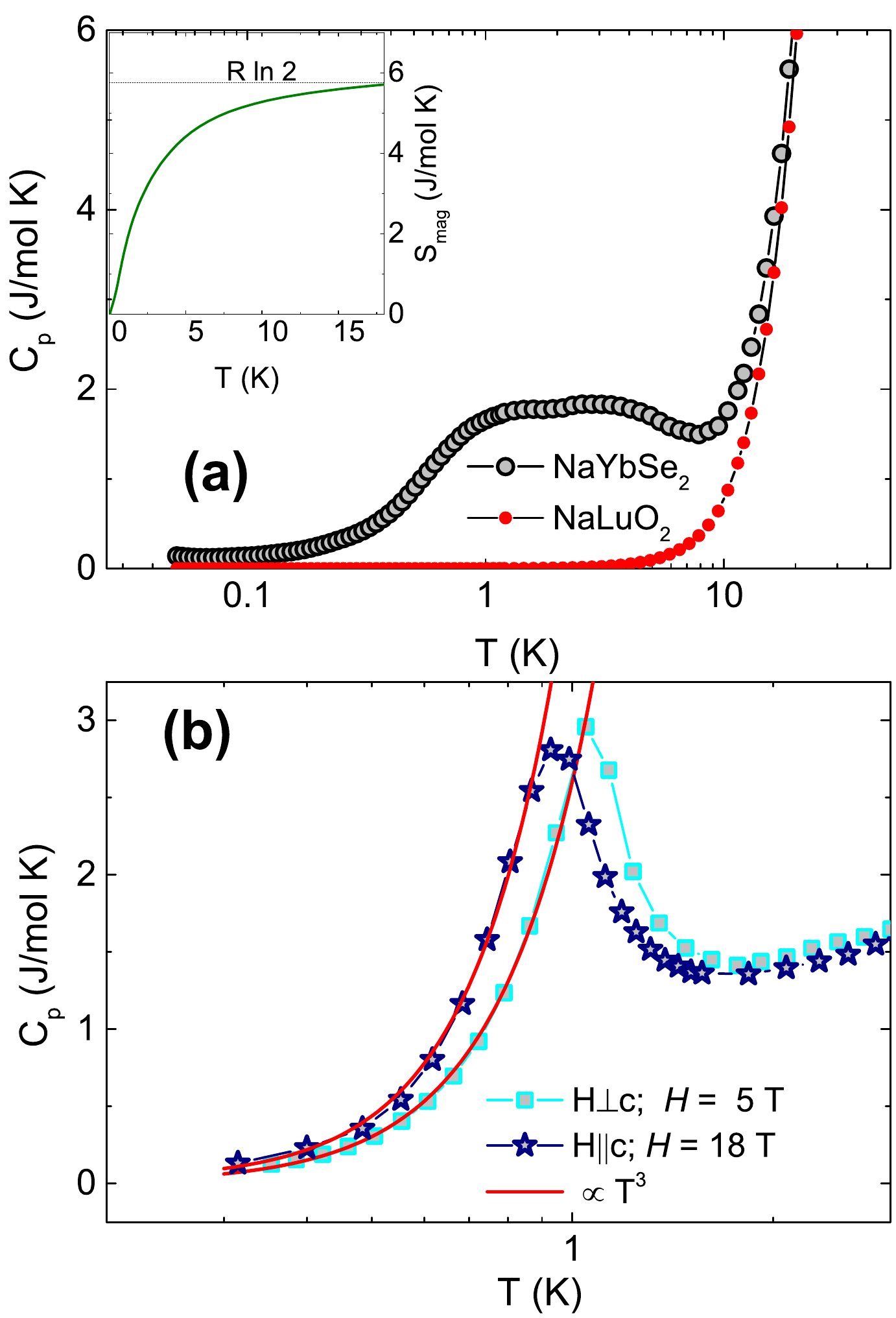}
 \caption{(a) Zero-field specific heat of NaYbSe$_2$ and NaLuO$_2$. Inset: The calculated magnetic entropy as a function of temperature. The dotted line indicates the  value of $R$ln2.
 (b) $C_{\rm p}(T)$ data in the ordered states for $H\perp c$ (5~T) and $H\parallel c$ (18~T), together with a $T^3$ power law (see text).} \label{fig3}
\end{figure}

Figure~\ref{fig3}(a) shows the zero-field specific heat $C_{\rm p}(T)$ of NaYbSe$_2$ along with that of the non-magnetic reference NaLuO$_2$. First, it is evident that there is no magnetic LRO down to $50\,\rm mK$ in NaYbSe$_2$,  which indeed points to a possible QSL ground state, similar to NaYbO$_2$ and NaYbS$_2$~\cite{baenitz2018,Ranjith2019}. Second, we observed a superposition of two broad maxima at low temperatures, $T_l\approx1.1\,\rm K$ and $T_h\approx2.75\,\rm K$. The zero-field $C_{\rm p}(T)$ of NaYbSe$_2$ contains several contributions and can be written as $C_p(T)=C_\text{nuc}(T)+C_\text{mag}(T)+C_\text{latt}(T)$.
A sharp upturn towards low temperatures~\cite{Note1} results from the high-temperature tail of a Schottky-type nuclear contribution $C_{\rm nuc}(T)=\alpha/T^2$. The magnetic contribution $C_{\rm mag}(T)$ to the specific heat is obtained by correcting for the nuclear contribution $C_\text{nuc}(T)$ and subtracting the lattice contribution $C_\text{latt}(T)$, which is estimated from the specific heat of the isostructural reference compound NaLuO$_2$ after correcting for the mass difference.
Towards low temperatures, $C_\text{mag}$ shows a nearly linear temperature dependence with a coefficient $\gamma \simeq$ 1~J/mol K$^2$~\cite{Note1}, which clearly evidences a gapless QSL ground state. The calculated magnetic entropy $S_{\rm mag}(T)\left( = \int_{0.05~K}^{T}\frac{C_{\rm{mag}}(T')}{T'}dT'\right)$ is plotted in the inset of Fig.~\ref{fig3}(a). At $T\approx15\,\rm K$, it reaches  $R\ln2$, as expected for an effective spin-$\frac{1}{2}$ state.

\subsection{Field-induced effects:}

\subsubsection{Magnetization studies}

The isothermal magnetization [Fig.~\ref{fig2}(b)] measured at $T=0.47\,\rm K$ for $H\bot c$ shows a clear plateau between 3 and 5~T at about one third of the saturation magnetization $M^\text s_{\perp}\simeq1.5~\mu_{\rm B}$, when substracting the  van Vleck contribution. The $M_\text s/3$ plateau is a manifestation of an up-up-down (uud) bound state spin arrangement, which is predicted by the mean field theory~\cite{chubokov1991}. Furthermore, the observation of the $M_\text s/3$ plateau for $H\bot c$ and the linear rise of the $M(H)$ curves for $H\|c$ are typical for an easy-plane type of magnetic anisotropy. In contrast, the linear behaviour of $M(H)$ for $H\|c$ suggests that the threefold rotational
symmetry is not broken by the applied field, rather a three-sublattice umbrella-like state emerges with the spins gradually canting out of the $ab$ plane. Correspondingly, no plateau in $M(H)$ was observed for $H\|c$.

To learn more about the field-induced effects, we have measured $\chi(T)$ at higher  fields for both $H\bot c$ and $H\|c$~\cite{Note1}. Above 2~T, a kink in $\chi(T,H)$ was observed at low temperatures for $H\bot c$, which confirms the field-induced magnetic ordering in NaYbSe$_2$ above $\mu_0H_\bot\approx2\,\rm T$. No such anomaly was observed for $H\|c$ up to $\mu_0H_\|\approx7\,\rm T$.

\subsubsection{Specific heat studies}

\begin{figure}
  \centering
  \includegraphics[clip,width=1\columnwidth]{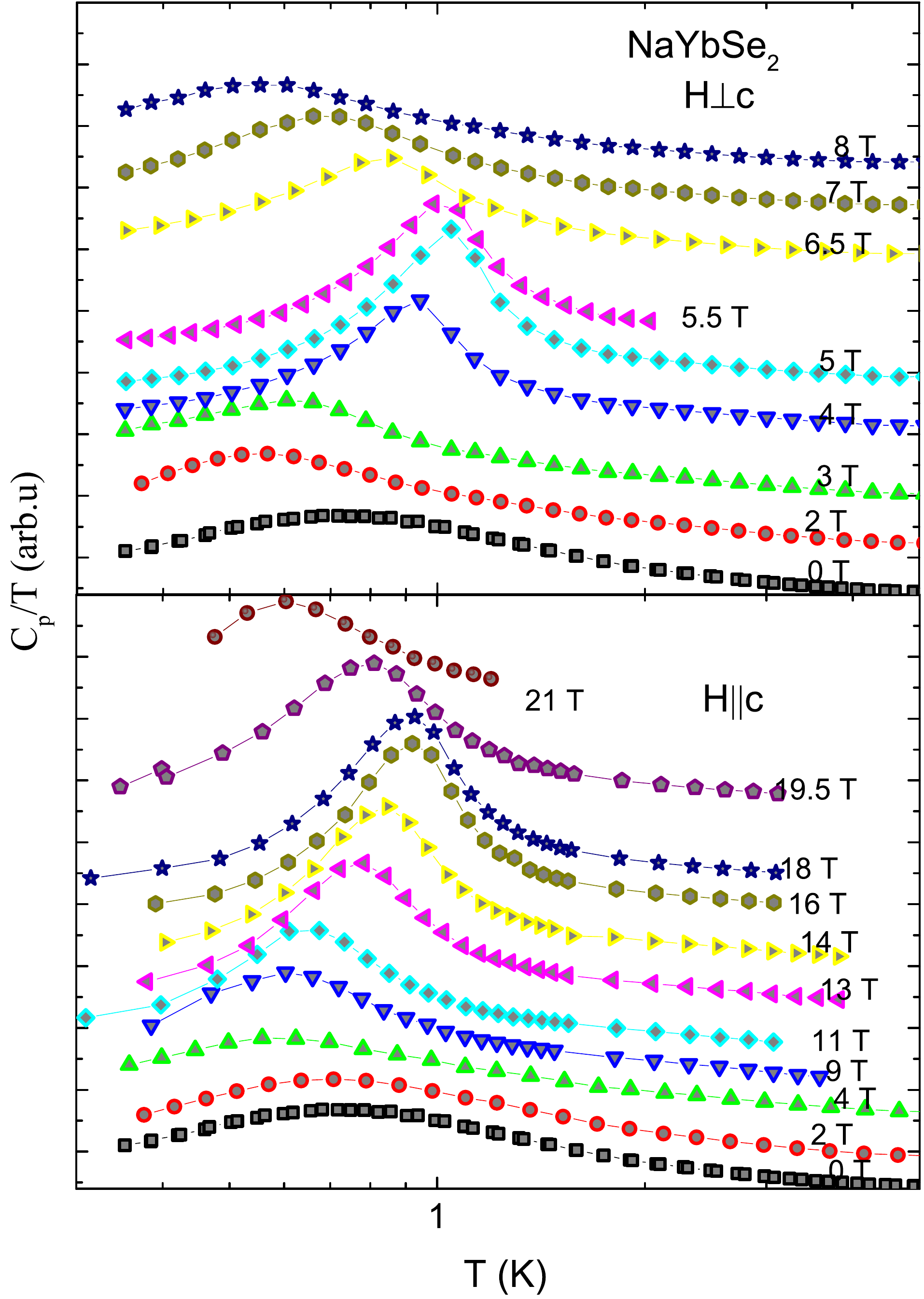}
   \caption{Temperature-dependent specific heat of NaYbSe$_2$, measured at different applied fields with orientations as indicated.}\label{fig4}
\end{figure}

$C_p(T)$ data measured at different applied fields, clearly evidencing the field-induced magnetic transitions in both directions, are shown in Fig.~\ref{fig4}. The order manifests as a peak that evolves from the low-$T$ regime upon applying the magnetic fields. An external field of $2\,\rm T$ in the $ab$ plane leads to a magnetic transition at $T_{\rm N}\simeq0.5\,\rm K$, whereas no such transition was observed for fields below $2\,\rm T$. As shown in Fig.~\ref{fig4}(a), this field-induced magnetic transition shifts towards higher temperatures with increasing magnetic field, reaching a maximum transition temperature at $5\,\rm T$ for $H\perp c$. A further increase of field results in a smooth decrease of $T_{\rm N}$, which finally vanishes above $\mu_0H_\bot\approx9\,\rm T$. Therefore, a pronounced reentrant behaviour of $T_{\rm N}$ is observed which includes the uud phase for $H\bot c$.  No field-induced ordering was observed in this field range (2 to 8~T) for $H\|c$, whereas a field-induced transition with $T_{\rm N}\simeq0.5\,\rm K$ appeared at $\mu_0H_\|\approx9\,\rm T$. The ordering temperature $T_{\rm N}(H)$ follows a similar trend, i.\,e. a systematic initial increase of $T_{\rm N}$ up to $0.95\,\rm K$ at $16\,\rm T$, and a shift towards lower temperatures with further increase of field up to $\mu_0H_\|=21\,\rm T$. As seen in Fig.~\ref{fig3}(b),  $C_p(T)$ below $T_{\rm N}$  (at $5\,\rm T$  for $H\bot c$  and $18\,\rm T$ for $H\| c$) follows a well defined $T^3$ behaviour, which could be assigned to the emergence of 3D magnons in a long-range ordered antiferromagnetic state.

\subsubsection{NMR measurements}

\begin{figure}
  \centering
  \includegraphics[clip,width=1\columnwidth]{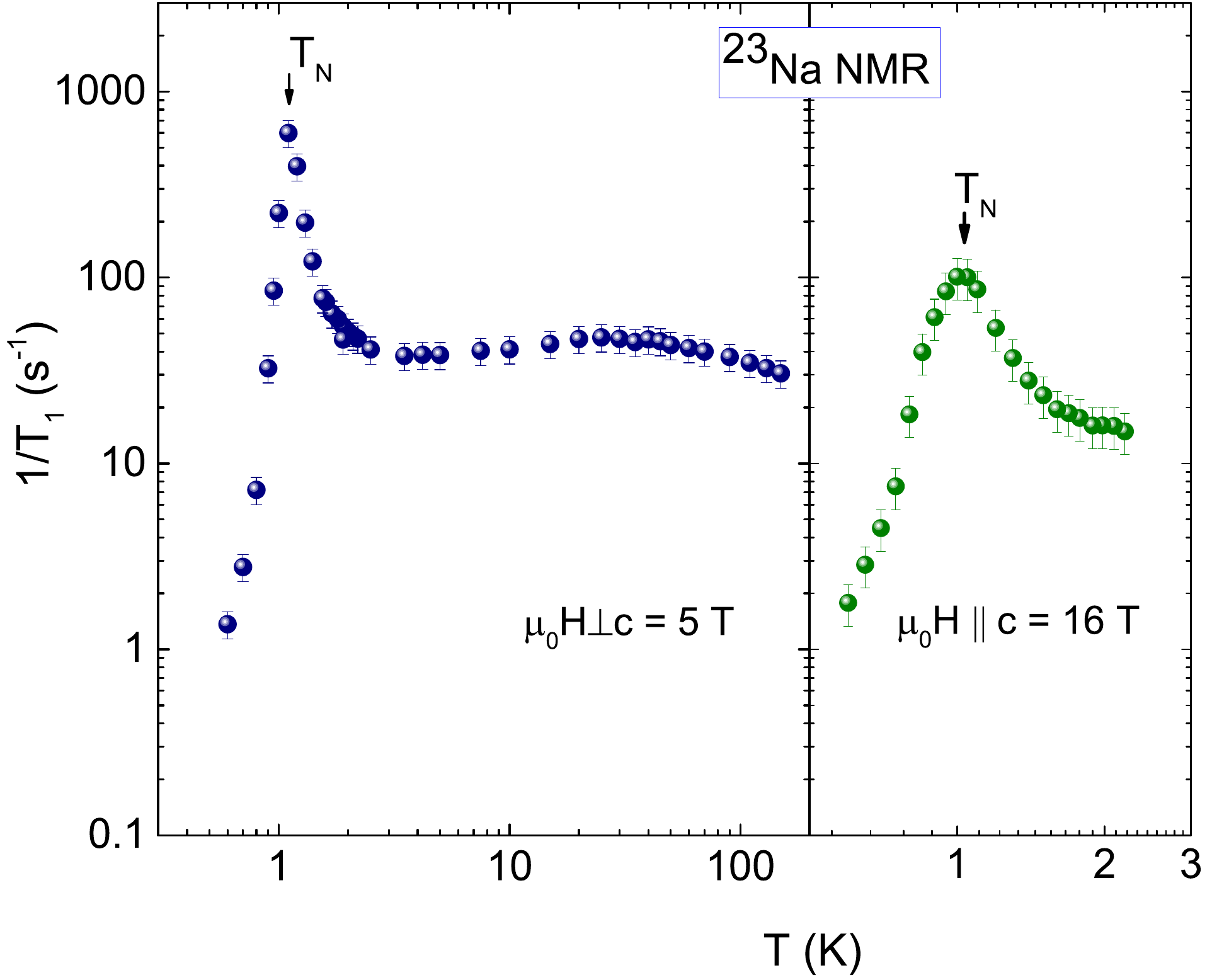}
  \caption{$^{23}$Na NMR spin-lattice relaxation rate in the magnetic ordered states for the field in plane at $\mu_0H_\perp  c$  = 5~T and for the fields out of plane at $\mu_0H_\parallel c$ = 16~T.} \label{fig6}
\end{figure}
We used $^{23}$Na NMR measurements to further  probe the field-induced magnetic ordering~\cite{Note1}. At higher temperatures, well-resolved $^{23}$Na NMR spectra are observed, consisting of a sharp central line and two satellites, as expected for a spin-$\frac32$ nucleus~\cite{Note1}. Upon decreasing the temperature, the NMR spectra are monotonously broadening, and show an abrupt increase of inhomogeneous broadening below 1.1 K~\cite{Note1}.   The sudden broadening of the NMR spectra  below 1.1~K is due to the field-induced magnetic  LRO in NaYbSe$_2$, where the spectral width is originating from the static internal field distribution at the $^{23}$Na sites. The NMR spin-lattice relaxation rate  NMR $1/T_1$ probes the low-energy excitations, which clearly evidences the field induced magnetic ordering at elevated fields. As seen in Fig.~\ref{fig6}, 1/$T_1$  measured for the fields in plane at  $\mu_0H_\bot\approx5\,\rm T$ and for the fields out of plane at $\mu_0H_\parallel\approx16\,\rm T$ show a clear peak at around  $T_{\rm N}\simeq$1~K corresponding to the field-induced magnetic LRO in both field directions.
\section*{Discussion}

The eightfold degenerate $j=7/2$ states of the Yb$^{3+}$ ions are split into four Kramer's doublets. The ground state doublet is well separated from the first excited doublet by an energy gap $\Delta/k_{\rm B}\approx160\pm30$~K. Therefore, the low-temperature properties can be well described by parameterizing the ground state with a pseudospin-$\frac{1}{2}$ and corresponding effective $g$ factors. The R\=3m space group symmetry allows for a NN exchange Hamiltonian, including the Zeeman coupling to the external field $\mu_0H$, of the general form

\begin{equation}
\begin{aligned}
	{\cal H}
	&=
	\sum_{\left\langle ij\right\rangle}\left\{
	J_\perp\left(S_i^xS_j^x+S_i^yS_j^y\right)
	+J_zS_i^zS_j^z
	\right.\\&\phantom{=\sum_{\left\langle ij\right\rangle}}\left.
	+\frac12J_\Delta\left(
	{\rm e}^{{\rm i}\phi_{ij}}S_i^-S_j^-
	+{\rm e}^{-{\rm i}\phi_{ij}}S_i^+S_j^+
	\right)
	\right.\\&\phantom{=\sum_{\left\langle ij\right\rangle}}\left.
	+\frac1{2\rm i}J_{yz}\left[
	{\rm e}^{{\rm i}\phi_{ij}}\left(S_i^zS_j^++S_i^+S_j^z\right)
	\right.\right.\\&
	\phantom{=\sum_{\left\langle ij\right\rangle}
	\frac1{2\rm i}J_{yz}}\left.\left.
	-{\rm e}^{-{\rm i}\phi_{ij}}\left(S_i^zS_j^-+S_i^-S_j^z\right)
	\right]
	\right\}
	\\&\phantom{=}
	-\mu_0\mu_\text B\sum_i\left[
	g_\perp\left(H_xS_i^x+H_yS_i^y\right)
	+g_\parallel H_zS_i^z
	\right],
	\\
	\phi_{ij}&=
	\left\{
	\begin{matrix}
	0,&\vec R_i-\vec R_j=(\pm1,0)\\
	\frac{2\pi}3,&
	\vec R_i-\vec R_j=\pm\left(-\frac12,\frac{\sqrt3}2\right)\\
	-\frac{2\pi}3,&
	\vec R_i-\vec R_j=\pm\left(-\frac12,-\frac{\sqrt3}2\right)
	\end{matrix}
	\right.
\end{aligned}
\label{eqHam}
\end{equation}

with $S_\ell^{\pm}=S_\ell^x\pm{\rm i}S_\ell^y$. Here,  $J_\perp=\frac12(J_x+J_y)$ is the rotationally invariant exchange in the $ab$ plane, and $J_z$ is the exchange component parallel to the $c$ axis~\cite{Note1}. The symmetry-allowed directional-dependent parts are $J_\Delta=\frac12(J_x-J_y)$ and $J_{yz}$, the latter describing the off-diagonal exchange coupling between spin components in the plane perpendicular to the bond direction $\vec R_i-\vec R_j$. We note that $J_\Delta$ and $J_{yz}$ are the energies for a $|\Delta S|=2$ and $|\Delta S|=1$ excitation, respectively, and we expect these to be small compared to $J_z$ and $J_\perp$. We note that in NaYbSe$_2$, the Yb ions reside on highly symmetric sites which precludes the antisymmetric (Dzyaloshinsky-Moriya) interaction for this compound~\cite{moriya:60}.

\subsection{Low-field thermodynamics:}

From the above spin Hamiltonian, the Curie-Weiss temperatures can be expressed  as $\Theta_{\rm CW}^{\bot} = -(3/2)J_{\bot}/k_\text B$ and $\Theta_{\rm CW}^{\|}=-(3/2)J_{z}/k_\text B$ for the field in the $ab$ plane and parallel to the $c$ axis, respectively~\cite{schmidt:17}. Our low temperature CW analysis ($\Theta_{\rm CW}^{\bot}\simeq$  -7~K and $\Theta_{\rm CW}^{\|}\simeq$ -3.5 ) yields $J_{\perp}/k_\text B \simeq$ 4.7~K and $J_z/k_\text B \simeq$ 2.33~K, which indicates a less anisotropic exchange compared to NaYbS$_2$~\cite{baenitz2018}.

It is worth noting that the presence of broad maxima in the susceptibility and specific heat is consistent with a quasi-two-dimensional scenario. For an isotropic $S=1/2$ TLAF with exchange $J$, such a broad maximum of $\chi(T)$ is expected at $k_\text BT_{\rm max}^{\chi}\lesssim J$~\cite{schmidt:17}. Below $70\,\text K$, the overall behaviour of both $\chi_\bot(T)$ and $\chi_\parallel (T)$ could be well described by the following Pad\'{e} approximation to the high-temperature series expansion given by~\cite{elstne1993},
\begin{equation}\label{Chi-equ}
   \chi(T)
   =
   \left.\frac{N_\text L\mu_0g^2\mu_{\rm B}^2}{4k_\text BT}
   \frac{1+b_1x+...+b_6x^6}{1+c_1x+...+c_7x^7}\right|_{x=J/(4k_\text BT)}.
\end{equation}
The coefficients $\{b_i\}$ and $\{c_i\}$ are listed in Ref.~\cite{Tamura2002}.

As seen in Fig.~\ref{fig2}(a), the $\chi(T)$ data below $70\,\rm K$ can be fitted well by Eq.~(\ref{Chi-equ}). Above $70\,\rm K$, a deviation was observed, which is expected due to the thermal population of the energetically higher CEF doublets. The obtained fitting parameters, $g_{\perp}\simeq3.0(2)$, $J_{\perp}/k_{\rm B}\simeq4.5(3)\,\rm K$ and $g_{\parallel}\simeq1.2(1)$, $J_{z}/k_{\rm B}\simeq2.4(3)\,\rm K$ are in good agreement with our ESR $g$-factors ($g_{\perp}\approx3.1$ and $g_{\parallel}\simeq1.2$) and the exchange couplings ($J_{\perp}/k_\text B\approx4.7\,{\rm K}$ and $J_{z}/k_{\rm B}\simeq2.33\,\rm K$), derived from the low temperature Curie-Weiss analysis. From Eq.~(\ref{eqHam}), the expression for the saturation field can be derived as $\mu_0H_\perp^{\rm sat}=\frac{9SJ_\perp k_{\rm B}}{\mu_{\rm B}g_\perp}$ and $\mu_0H_\parallel^{\rm sat}=\frac{3S(2J_z+J_\perp) k_{\rm B}}{\mu_{\rm B}g_\parallel}$ for the field in the $ab$-plane
and parallel to the $c$-axis, respectively~\cite{Note1}. The calculated values  $\mu_0H_\perp^{\rm sat}\simeq$ 10.2~T and $\mu_0H_\parallel^{\rm sat}\simeq$ 21~T are slightly below the observed values of 12 and 25~T, respectively. This might  indicate the presence of antiferromagnetic next-NN interactions in NaYbSe$_2$.

Moreover, the zero field $C_{\rm mag}(T)$ shows a superposition of two broad peaks at around 1.1 and 2.75~K~\cite{Note1}. This is rather unusual for a 2D nearest-neighbor TLAF, for which one would expect only a single maximum at $k_\text BT\approx 0.55J_{\rm 2D}$~\cite{elstne1993}. However, such double broad maxima in $C_{\rm p}(T)$ are predicted for two-dimensional (2D) triangular/Kagom\'{e} lattice  antiferromagnets with a fully frustrated disordered ground state~\cite{Ishida1997,Elser1989}. A similar feature was experimentally observed in the recently reported 2D TLAF compound  Ba$_8$CoNb$_6$O$_{24}$~\cite{Rawl2017,Cui2018}. Recent numerical simulations  based on an exponential tensor renormalization group method reveal two crossover temperature scales in $C_{\rm mag}(T)$ for a spin-$\frac 12$ TLAF, at $T_l/J\simeq$ 0.2 and $T_h/J\simeq$ 0.55~\cite{Chen2019}. The anticipated positions of the maxima with $J_{\perp}\simeq$ 4.7~K closely match with our experimental data. The absolute values of the maxima are very sensitive to the degree of magnetic frustration. Broad maxima with $C_{\rm mag}^{\rm max}\simeq$  0.44~R and 0.35~R are expected for a non-frustrated square lattice and spin chain, respectively, where R is the gas constant~\cite{johnston2000,hofmann2003}. In case of a highly frustrated triangular-lattice, $C_{\rm mag}^{\rm max}$ is expected to be much smaller (0.22~R)~\cite{elstne1993,bernu2001}. Our value of $C_{\rm mag}^{\rm max}\simeq$ 0.218~R matches well with the one expected for a TLAF~\cite{Note1}.

\subsection{Field-induced magnetic order:}

The exchange anisotropy, which is common for an effective spin-$\frac{1}{2}$ ground state doublet of Yb$^{3+}$, is estimated as $J_z/J_\perp\simeq0.5$ for NaYbSe$_2$. In TLAFs with isotropic exchange as well as with an easy-plane anisotropy, the ground state is a $120^{\circ}$ coplanar state. In contrast, multiple ordered phases (low temperature $120^\circ$ state and an intermediate temperature collinear state) are predicted for an easy-axis type anisotropy~\cite{Miyashita1986,Matsubara1982,Melchy2009,ranjith2016,ranjith2017}. NaYbSe$_2$ does not show any magnetic LRO at zero fields. Instead, we observed a gapless and strongly fluctuating state. Only towards higher applied fields, the magnetic order is induced and stabilized with increasing field strength. Recently, Z.~Zhu et al. constructed a four-dimensional extension of the phase diagram for the anisotropic bond-dependent XXZ model with a next-NN coupling $J_2$, which can host a spin liquid phase and is continuously connected to a  spin-liquid phase of the isotropic $J_1\mbox{-}J_2$ model~\cite{zhu2018}. Alternatively to the impact of next-NN couplings on the ground state, it should be noted that the off-diagonal terms in the bond exchange matrices can influence the ground state and yield strong zero-field fluctuations. So far, we do not have any estimate for these higher exchange terms. In the case of YbMgGaO$_4$, these interactions are reported to be rather small~\cite{zhang2018}.

\begin{figure}
  \centering
  \includegraphics[clip,width=1\columnwidth]{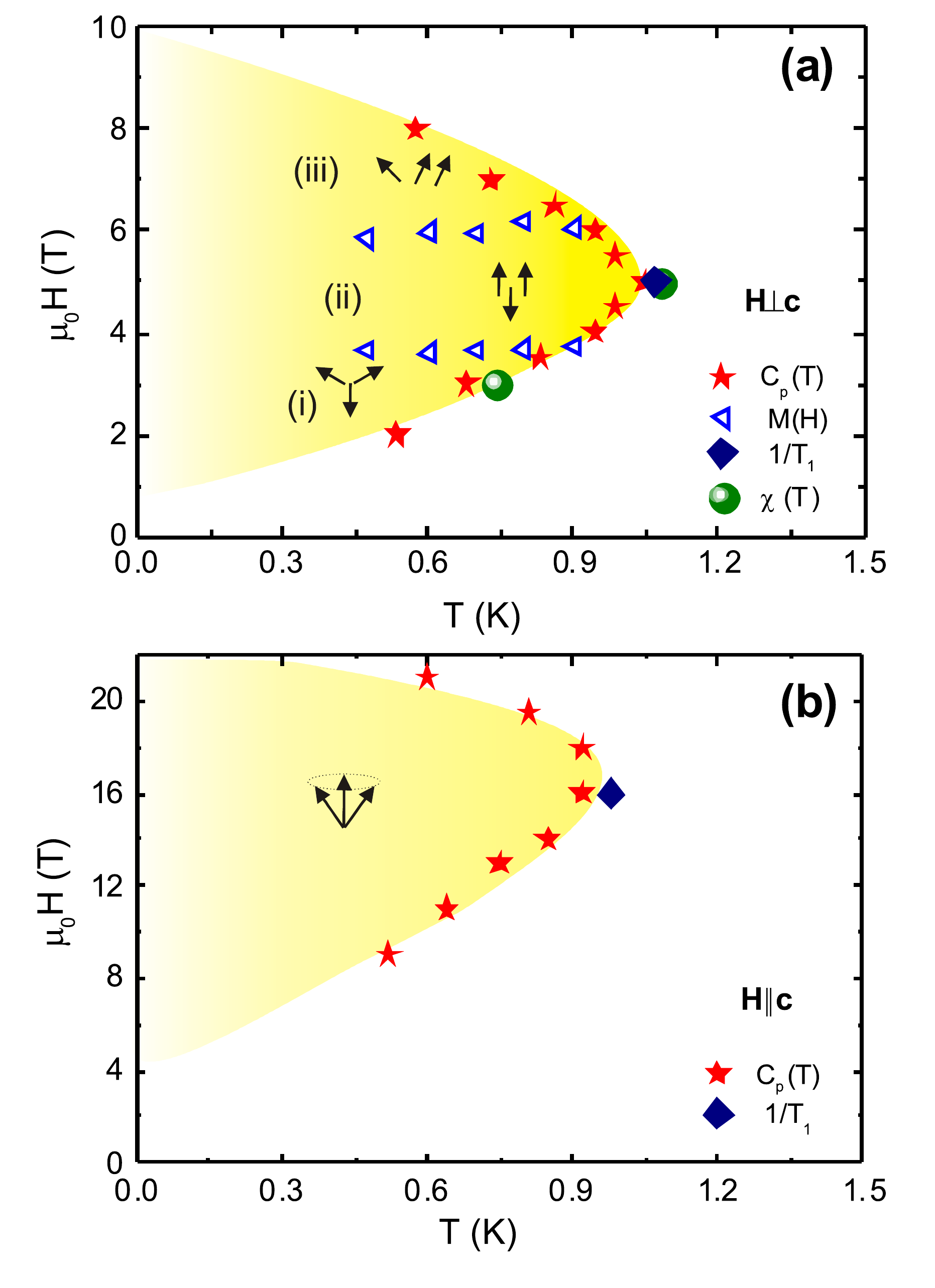}
  \caption{Field-temperature phase diagram of NaYbSe$_2$ for $H\bot c$ and $H\|c$.} \label{fig5}
\end{figure}

 Figures~\ref{fig5}(a) and ~\ref{fig5}(b) represent the $H\mbox{-}T$ phase diagrams for both $H\perp c$ and $H\parallel c$ directions, respectively, constructed from the field dependent $C_{\rm p}(T)$ and temperature dependent $M(H)$ measurements~\cite{Note1}. For $H\perp c$, we may assign different spin configurations: In the low-field regime (below 1~T), a highly fluctuating QSL-like phase with persistent gapless excitations was observed instead of the 120$^{\circ}$ ordered phase expected for a TLAF with easy-plane anisotropy. In region-$(i)$, above approximately 1~T, an oblique 120$^{\circ}$ spin structure ($Y$-coplanar) structure may be stabilized by the application of a moderate external field along the $ab$ plane. Upon increasing the field, the transition from $Y$-coplanar to the collinear uud phase appears at about $\approx$3~T [region ($ii$)], which manifests as the $M_\bot^\text s/3$ plateau in $M(H)$ [see fig.~\ref{fig2}(b)]. With further increasing field, a 2:1-coplanar magnetic phase might appear above $\mu_0H_\bot\approx6\,\rm T$ [region ($iii$)], before reaching full saturation at $\mu_0H_\bot^s\approx12\,\rm T$. It should be mentioned that the phase boundaries from the specific heat represent second order phase transitions, whereas the open symbols obtained from $M(H)$ represent cross-over lines between the different spin textures.

A magnetic field parallel to the $c$-axis does not break the threefold rotational symmetry of the crystal. Therefore, a single magnetically ordered umbrella-shaped three-sublattice phase is induced, here at fields $\mu_0H_\|\gtrsim9\,\rm T$, substantially higher than the in-plane field. In concordance with this, the saturation field at low temperatures exceeds $\mu_0H_\|\simeq21\,\rm T$, where $T_\text N$ is still finite.
Single-crystal neutron diffraction measurements under external magnetic fields will be necessary to determine the magnetic structure in more detail.

The overall shape of the phase diagram is in good agreement with results from classical Monte-Carlo simulations~\cite{Seabra2011}, apart from the low-field disordered regime. Similar $H\mbox{-}T$ phase diagrams for $H\perp c$ and $H\parallel c$ are also reported for the effective spin-$\frac{1}{2}$ TLAF Ba$_3$CoSb$_2$O$_9$~\cite{Quirion2015,Koutroulakis2015}. In contrast to these, the field-induced magnetic transitions are highly anisotropic in NaYbSe$_2$. Moreover, the $H\mbox{-}T$ phase diagram contains both a QSL phase and a full range of ordered phases for a prototypical $S=1/2$ TLAF, again different to Ba$_3$CoSb$_2$O$_9$ or any other TLAFs. In this sense, NaYbSe$_2$ provides a unique opportunity to study the low temperature properties of pseudospin $S=1/2$ triangular-lattice antiferromagnets.

\section*{Conclusion}

In conclusion, we have synthesized and investigated high quality single crystals of NaYbSe$_2$, a Yb-based triangular-lattice QSL candidate without any inherent disorder. The ground state doublet is well separated from the first excited state doublet, inducing a (pseudo)spin-$\frac{1}{2}$ ground state, as confirmed by  ESR measurements and a low temperature entropy analysis. Zero-field measurements confirm the absence of magnetic LRO down to 50~mK, suggesting a gapless QSL ground state. Strong bond frustration introduces a melting of the predicted 120$^{\circ}$ order at zero-fields. The QSL ground state is found to be unstable against applied magnetic fields. A clear $\frac{M_s}{3}$ magnetization plateau, associated with an up-up-down spin configuration, was observed for the $H\bot c$ direction. The reentrant behaviour of $T_{\rm N}$, originating from the thermal and quantum spin fluctuations, is observed for both directions. The obtained $H\mbox{-}T$ phase diagram is highly anisotropic and includes different quantum phases expected for a triangular-lattice antiferromagnet, which classifies NaYbSe$_2$ as a unique model system and opens up future theoretical and experimental studies.
\section*{Acknowledgment}
We thank A. Mackenzie, H. Rosner, C. Geibel, L. Hozoi, J. Rau and A. L. Chernyshev for fruitful discussions. We thank C. Klausnitzer and H. Rave for technical support. We acknowledge the support from the Deutsche Forschungsgemeinschaft (DFG) through SFB 1143, and the W\"{u}rzburg-Dresden Cluster of Excellence on Complexity and Topology in Quantum Matter - ct.qmat (EXC 2147, project-id 39085490), as well as from the HLD at HZDR, member of the European Magnetic Field Laboratory (EMFL). AMS thanks the SA-NRF (105702) for financial assistance.

\end{document}